\documentclass[fleqn,10pt]{wlscirep}
\usepackage{amsmath,amsfonts,amssymb,graphics,graphicx,epsfig,color,times,bbm,subfigure,hyperref,keyval}

\title{Generating multi-atom entangled W states via light-matter interface based fusion mechanism}
\author[1,2]{Xue-Ping Zang}

\author[1,$^\ast$]{Ming Yang}

\author[3,$^\dag$]{Fatih Ozaydin}

\author[4]{Wei Song}

\author[4]{Zhuo-Liang Cao}

\affil[1]{School of Physics {\&} Material Science, Anhui University, Hefei 230601, People's Republic of China}
\affil[2]{Department of Mechanical and Electronic Engineering, Chizhou University, Chizhou 247000, People's Republic of China}
\affil[3]{Department of Information Technologies, Isik University, Sile, Istanbul, 34980, Turkey}
\affil[4]{Institute for Quantum Control and Quantum Information, School of Electronic and Information Engineering, Hefei Normal University, Hefei 230601,  People's Republic of China}
\affil[$^\ast$]{mingyang@ahu.edu.cn}
\affil[$^\dag$]{mansursah@gmail.com}

\begin{abstract}
W state is a key resource in quantum communication. Fusion technology has been proven to be a good candidate for preparing a large-size W state from two or more small-size W states in linear optical system\cite{ozdemir}. It is of great importance to study how to fuse W states via light-matter interface. Here we show that it is possible to prepare large-size W-state networks using a fusion mechanism in cavity QED system. The detuned interaction between three atoms and a vacuum cavity mode constitute the main fusion mechanism, based on which two or three small-size atomic W states can be fused into a larger-size W state. If no excitation is detected from those three atoms, the remaining atoms are still in the product of two or three new W states, which can be re-fused. The complicated Fredkin gate used in the previous fusion schemes is avoided here. W states of size $2$ can be fused as well. The feasibility analysis shows that our fusion processes maybe implementable with the current technology. Our results demonstrate how the light-matter interaction based fusion mechanism can be realized, and may become the starting point for the fusion of multipartite entanglement in cavity QED system.
\end{abstract}

\begin{document}
\flushbottom
\maketitle

\section*{Introduction}
Quantum entanglement is a crucial resource in many quantum information and quantum communication tasks. Recently, quantum entanglement has attracted more and more attention due to its various fundamental quantum features. The non-locality \cite{bell} is a typical feature of entanglement and it brings many applications in implementing various quantum information processing schemes, for instance, quantum dense coding \cite{mattle}, quantum key distribution \cite{gisin}, quantum cryptography \cite{ekert, bennett1, nielsen}, etc. Thus, to prepare entangled states between different particles is essential for demonstrating quantum non-locality. Bipartite entanglement is the simplest form of entanglement, whose properties have been well understood so far. Nevertheless, the structure and properties of multipartite entanglement are far from being clear. Moveover, it was pointed out that the multipartite entanglement contains stronger non-locality, which offers significant advantages for quantum information and computation tasks.

A remarkable difference between bipartite entanglement and multipartite one is the classification of them. Greenberger-Horne-Zeilinger (GHZ) state \cite{greenberger}, W state \cite{dur} and cluster state \cite{briegel} are three typical classes of multipartite entangled states. Recently, researches have shown that for different quantum information processing tasks, the corresponding multipartite entangled states are needed. For instance, cluster states are basic resources for measurement-based quantum computation \cite{raussendorf}, GHZ states are the best quantum channels for teleportation \cite{zhao} and quantum key distribution \cite{kempe}, and W states are required for secure quantum communication \cite{wang,liu}. These different classes of multipartite entangled states cannot be converted into each with local operations and classical communication \cite{dur00}. Among them, $W$ state is a special class of multipartite entangled state, which possesses many particular properties. For instance, the entanglement of $W$ state is robust against disposal of particles\cite{dur}. D'Hondt \emph{et al} showed that W state plays an important role in the leader election problem in anonymous quantum networks \cite{hondt}. Ozaydin studied quantum Fisher information (QFI) of W state in the basic decoherence channels \cite{ozaydin14, ozaydin15}. C. Dag et al. showed that the quantum coherence of W states enable high efficiency in quantum thermalization of a single mode cavity \cite{ceren1}. Three-qubit W and GHZ states have been recently demonstrated on an NMR quantum information processor \cite{arvind1}, and the equivalence of superpositions of W states to GHZ states under local filtration has been studied \cite{arvind2}. In addition, self-testing (a device-independent method for determining the nature of a physical system or device) of W state is totally different from self-testing of GHZ and cluster states \cite{wu,pal}.

It is worth mentioning that an arbitrary bipartite state of qubits can be prepared from a maximally entangled two-qubit state via local operations and classical communication (LOCC). However, the same situation cannot happen in the case of multipartite entangled states because of the inequivalent relations between different classes of multipartite entangled states, and the local conversion between $W$ state and $GHZ$ state only can be done in an approximate way \cite{walther,yu}. Therefore, simple and efficient schemes to prepare large-size multipartite entangled states are of great importance.

Recently, quantum state fusion and expansion technology have been presented and experimentally demonstrated as efficient ways for creating large-size multipartite entangled states \cite{ozdemir,zeilinger,browne,tashima,walther,tashima4, tashima2, ikuta,ozaydin,yesilyurt,bugu}. Fusion technology can create a multipartite entangled state of larger number of particles by fusing two or more multipartite entangled states of smaller number of particles, with the condition that the access is approved only to one qubit of each of the entangled states. On the other hand, with expansion technology, the number of qubits of the original entangled state is expanded by one or two each time. Expansion and preparation of GHZ state and cluster state are well known \cite{zeilinger,browne}, but it will be another story for expanding or fusing W state because of the inequivalent relation between these two types of entangled states. T. Tashima \emph{et al.} showed that two EPR photon states can be merged into a three-photon W state using LOCC \cite{tashima}. In addition, to expand a polarization entangled $|W_N\rangle$ state to a $|W_{N + n}\rangle$ state, T. Tashima \emph{et al.} designed several efficient methods\cite{tashima4, tashima2, ikuta}. S. K. \"{O}zdemir \emph{et al.} proposed a scheme for fusing two polarization entangled states, $|W_N\rangle$ and $|W_M\rangle$, to a larger-size entangled state $|W_{N+M-2}\rangle$ \cite{ozdemir}.

The previous fusion schemes for W states are limited by some constraints, which will inevitably decrease the fusion efficiency. For instance, S. K. \"{O}zdemir's scheme for fusing two polarization entangled states $|W_N\rangle$ and $|W_M\rangle$ to a larger-size entangled $|W_{N+M-2}\rangle$ state only works for $N, M>2$ cases\cite{ozdemir}. To relax this constraint, F. Ozaydin \emph{et al.} proposed several new fusion schemes such that Bell states, i.e. W states of sizes $N=2, M=2$ can be fused as well. Therefore the requirement of creating initial $W_3$ states has disappeared\cite{ozaydin,yesilyurt,bugu}. In addition, these schemes can fuse two, three or four W states of arbitrary size via accessing only one qubit of each W state through Fredkin gate, but the realization of Fredkin gate is not an easy task in experiment. To overcome this difficulty, Diker \emph{et al.} showed that in the fusion mechanisms, the Fredkin gate can be replaced with a CNOT gate and a Toffoli gate \cite{diker1} which can be efficiently implemented by measurement-based one way quantum computation, using weighted graph states \cite{TameWeighted}. On the other hand, since the efficiencies of fusion and expansion methods depend on the size of the initial W state to be used as the primary resource, Yesilyurt \emph{et al.} proposed a scheme for deterministic preparation of $W_4$ states \cite{yesilyurtActa}. Very recently, Han \emph{et al.} proposed a scheme for fusing an $n$-qubit $W$ state and an $m$-qubit $W$ state to an $(n+m-1)$-qubit W state without any ancillary photons with success probability $(n+m-1)/nm$, which is mainly based on the cross-Kerr nonlinearities \cite{han}. Li \emph{et al.}, based on cross-Kerr nonlinearities and a quantum dot coupled cavity system, proposed a similar fusion scheme too \cite{lina}. Although these two schemes can overcome the difficulty of Fredkin gate, they are still too much complicated to be realized in Lab.

 Although photon is the most promising flying qubit in quantum communication, the interface between flying and matter qubits is an equivalently important mechanism for quantum communication networks. For instance, a NOON state of photons in two superconducting resonators can be generated in circuit QED system \cite{SuSR}. Macroscopic arbitrary entangled coherent states (ECSs) of separate nitrogen-vacancy center ensembles (NVEs) can be generated via the coupling between NVEs and a superconducting flux qubit \cite{SongSR}. Teleportation of a Toffoli gate among three spatially separated electron spin qubits in optical microcavities can be made possible by using the coupling between electron spin and circularly-polarized photons \cite{HuSR}. A microwave photonic quantum bus is proposed for a strong direct coupling between the topological and conventional qubits \cite{XueSR}. The interface between the spin of a photon and an electron spin confined in a quantum dot embedded in a microcavity is applied to build universal ququart logic gates on the photon system with two freedoms \cite{LuoSR}. The controlled-NOT, Toffoli, and Fredkin gates between a flying photon qubit and diamond nitrogen-vacancy (NV) centers can be realized with the assistance of microcavities \cite{WeiSR}. The generation and reconstruction of arbitrary states can be realized in the ultrastrong coupling regime of light-matter interactions in Cavity QED system \cite{FelicettiSR}. So it is of great importance to study how to fuse the multipartite entangled states of flying and matter qubits via the interface of them. But, most of the above-mentioned expansion and fusion schemes of W states are only applicable to optical systems. Very recently, by using the interface between photons and electron spins confined in quantum dots embedded in a microcavity, Han \emph{et al.} proposed specialized fusion schemes for stationary electronic W states and flying photonic W states, respectively \cite{HanSR}. In addition, an N-qubit W state generation scheme has been proposed for N atoms trapped at separated quantum nodes (cavity-QED systems) by using linear optics \cite{fujii}. We have proposed a scheme for expanding atomic entangled W states by resonant interactions between atoms and cavity modes \cite{zang} and a scheme for fusing two atomic entangled W states to an entangled W state of a larger size in cavity QED system \cite{zang1}. In this paper, two new light-matter interaction based fusion mechanisms are proposed for generating large-size W-state networks in cavity QED system, where the complicated Fredkin gate is not needed, the requirement of creating initial $W_3$ states disappears too, and the interactions between atoms and the cavity mode are far-off-resonant, which make the proposed schemes feasible within the current technology.

\section*{Results}
\subsection*{ Fusion of Two W states}

\begin{figure}[t!]
\centering
\includegraphics[width=0.40\textwidth]{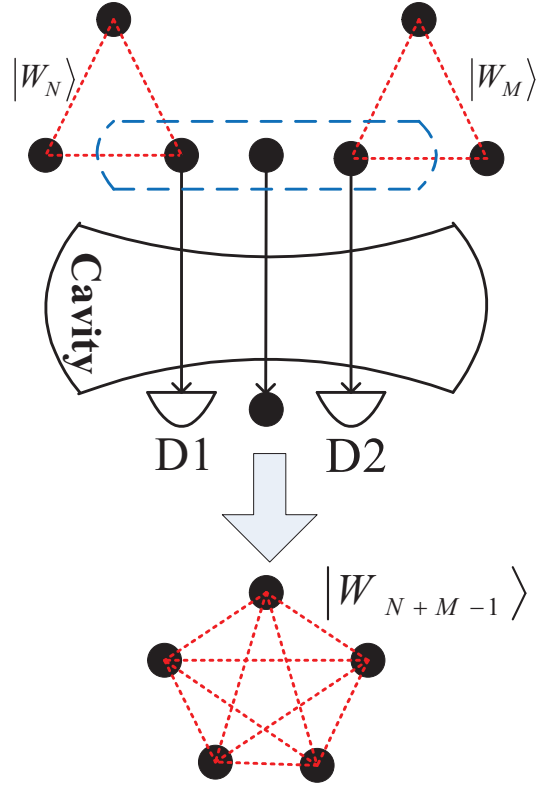}
\caption{The setup for fusion of  W states. $D1$ and $D2$ denote the atomic detectors. $N$-atom W state $|W_N\rangle$, $M$-atom W state $|W_M\rangle$ and an auxiliary atom are fused into a $(N+M-1)$-atom W state $|W_{N+M-1}\rangle$. }\label{Fig1}
\end{figure}

As depicted in FIG.\ref{Fig1}, to fuse $N$-partite W-state $|W_N\rangle_a $ and $M$-partite W-state $|W_M\rangle_b$ an auxiliary atom in ground state will be introduced. First, let's pick up one atom from each entangled system, and then send these two atoms and the auxiliary atom into the cavity and detect any two of the three atoms after flying out of the detuned cavity. If only one excitation is detected, the remaining $(N+M-1)$ atoms are successfully prepared in a $|W_{N+M-1}\rangle$ sate.
To simplify the descriptions of the fusion scheme, it is necessary to give the formulation of $N$-atom W state: $|W_N\rangle=|(N-1)_g,e\rangle/\sqrt{N}=[|(N-1)_g\rangle_{\widetilde{i}}\otimes |e_i\rangle+|(N-2)_g,e\rangle_{\widetilde{i}}\otimes |g_i\rangle]/\sqrt{N} $. Here $i$ marks the $i$th atom and the remaining $N-1$ atoms of $|W_N\rangle$ are marked by $\widetilde{i}$. $N-1$ atoms in the ground state $(|g\rangle)$ are described by $|(N-1)_g\rangle$, and all the possible combinations of $N-2$ atoms in the ground state $(|g\rangle)$ and one atom in the excited state $(|e\rangle)$ are described by $|(N-2)_g, e\rangle$.

To start the fusion process, the three atoms will be sent into the cavity. The far-off-resonant interaction between the cavity mode and the three atoms will lead the initial states
\begin{eqnarray}\label{eq1}
|W_N\rangle_a &=&|(N-1)_g,e\rangle/\sqrt{N} \nonumber\\
&=&[|(N-1)_g\rangle_{\widetilde{1}}\otimes |e_1\rangle+|(N-2)_g,e\rangle_{\widetilde{1}}\otimes |g_1\rangle]/\sqrt{N},
\end{eqnarray}
and
\begin{eqnarray}\label{eq2}
|W_M\rangle_b &=&|(M-1)_g,e\rangle/\sqrt{M} \nonumber\\
&=&[|(M-1)_g\rangle_{\widetilde{2}}\otimes |e_2\rangle+|(M-2)_g,e\rangle_{\widetilde{2}}\otimes |g_2\rangle]/\sqrt{M},
\end{eqnarray}
evolve to the following state (for detailed description of this interaction, see the discussion in the Methods section below):
\begin{equation}
\begin{aligned}
|W_N\rangle_a\otimes|W_M\rangle_b\otimes|g_3\rangle&=|(N-1)_g,e\rangle/\sqrt{N}\otimes|(M-1)_g,e\rangle/\sqrt{M}\otimes|g_3\rangle\\
&\longrightarrow 1/\sqrt{NM}|(N-1)_g\rangle_{\widetilde{1}}|(M-1)_g\rangle_{\widetilde{2}}e^{-i\lambda t}(A|g_1\rangle|e_2\rangle|e_3\rangle+A|e_1\rangle|g_2\rangle|e_3\rangle+B|e_1\rangle|e_2\rangle|g_3\rangle) \\
&+1/\sqrt{NM}|(N-1)_g\rangle_{\widetilde{1}}|(M-2)_g,e\rangle_{\widetilde{2}}( A|g_1\rangle|g_2\rangle|e_3\rangle+A|g_1\rangle|e_2\rangle|g_3\rangle+B|e_1\rangle|g_2\rangle|g_3\rangle)\\
&+1/\sqrt{NM}|(N-2)_g,e\rangle_{\widetilde{1}}|(M-1)_g\rangle_{\widetilde{2}}(A|g_1\rangle|g_2\rangle|e_3\rangle+B|g_1\rangle|e_2\rangle|g_3\rangle+A|e_1\rangle|g_2\rangle|g_3\rangle)\\
&+1/\sqrt{NM}|(N-2)_g,e\rangle_{\widetilde{1}}|(M-2)_g,e\rangle_{\widetilde{2}}|g_1\rangle|g_2\rangle|g_3\rangle,\label{eq3}
\end{aligned}
\end{equation}
where $A=(e^{-i3\lambda t}-1)/3, B=(e^{-i3\lambda t}+2)/3$.

After flying out of the cavity, any two (say $(1,2)$) of the three atoms will be detected. The detection result $|e_1\rangle|e_2\rangle$ means the failure of the fusion process. If the detection result is $|g_{1}\rangle|e_{2}\rangle$ or $|e_{1}\rangle|g_{2}\rangle$, the remaining atoms are in the following states
\begin{equation}\label{eq4}
|\varphi_{ge}\rangle=\frac{1}{\sqrt{NM}}[e^{-i\lambda t}A|(N-1)_g\rangle_{\widetilde{1}}|(M-1)_g\rangle_{\widetilde{2}}|e_3\rangle+A|(N-1)_g\rangle_{\widetilde{1}}|(M-2)_g,e\rangle_{\widetilde{2}}|g_3\rangle+B|(N-2)_g,e\rangle_{\widetilde{1}}|(M-1)_g\rangle_{\widetilde{2}}|g_3\rangle],
\end{equation}
or
\begin{equation}\label{eq5}
|\varphi_{eg}\rangle=\frac{1}{\sqrt{NM}}[e^{-i\lambda t}A|(N-1)_g\rangle_{\widetilde{1}}|(M-1)_g\rangle_{\widetilde{2}}|e_3\rangle+B|(N-1)_g\rangle_{\widetilde{1}}|(M-2)_g,e\rangle_{\widetilde{2}}|g_3\rangle+A|(N-2)_g,e\rangle_{\widetilde{1}}|(M-1)_g\rangle_{\widetilde{2}}|g_3\rangle],
\end{equation}
respectively. By controlling the velocity of the atoms appropriately, $\lambda t=2\pi/9$ can be satisfied, and thus the states in Eqs.(\ref{eq4},\ref{eq5}) will become:
\begin{equation}\label{eq6}
|\varphi_{ge}\rangle=\frac{1}{\sqrt{3NM}}[e^{-i2\pi/9}|(N-1)_g\rangle_{\widetilde{1}}|(M-1)_g\rangle_{\widetilde{2}}|e_3\rangle+|(N-1)_g\rangle_{\widetilde{1}}|(M-2)_g,e\rangle_{\widetilde{2}}|g_3\rangle+e^{i2\pi/3}|(N-2)_g,e\rangle_{\widetilde{1}}|(M-1)_g\rangle_{\widetilde{2}}|g_3\rangle],
\end{equation}
or
\begin{equation}\label{eq7}
|\varphi_{eg}\rangle=\frac{1}{\sqrt{3NM}}[e^{-i2\pi/9}|(N-1)_g\rangle_{\widetilde{1}}|(M-1)_g\rangle_{\widetilde{2}}|e_3\rangle+e^{i2\pi/3}|(N-1)_g\rangle_{\widetilde{1}}|(M-2)_g,e\rangle_{\widetilde{2}}|g_3\rangle+|(N-2)_g,e\rangle_{\widetilde{1}}|(M-1)_g\rangle_{\widetilde{2}}|g_3\rangle],
\end{equation}
where a global phase factor $e^{-i5\pi/6}$ has been discarded. Although the states in Eqs.(\ref{eq6},\ref{eq7}) are not standard W states:
\begin{eqnarray}\label{eq8}
|\varphi_{ge}\rangle&=&|\varphi_{eg}\rangle=\frac{\sqrt{N+M-1}}{\sqrt{3NM}}|W_{N+M-1}\rangle,
\end{eqnarray}
there are only relative phase differences between these two states and the standard W state, which can be removed by a classical pulse on any one of the $(N+M-1)$ atoms. The total success probability for the fusion process is
\begin{eqnarray}\label{eq9}
 P_{N+M-1}=\frac{2(N+M-1)}{3NM}.
\end{eqnarray}

If the detection result is $|g_1\rangle|g_2\rangle$, the third atom must be detected for continuing the scheme. If the third atom is detected in excited state, the remaining $N+M-2$ atoms will be left in W state.  If the third atom is detected in ground state, the state of the remaining $N+M-2$ atoms are factorized into the product of $|W_{N-1}\rangle\bigotimes|W_{M-1}\rangle$, which can be further fused by the same process.

\subsection*{ Fusion of Three W states}

\begin{figure}[t!]
\centering
\includegraphics[width=0.40\textwidth]{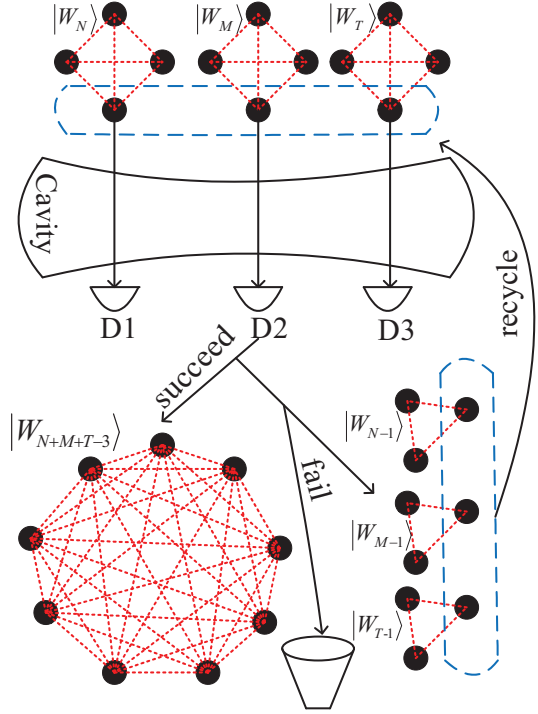}
\caption{The setup for fusion of  W states. $D1$, $D2$ and $D3$ denote the atomic detectors. $N$-atom W state $|W_N\rangle$, $M$-atom W state $|W_M\rangle$ and $T$-atom W state $|W_T\rangle$ are fused into a $(N+M+T-3)$-atom W state $|W_{N+M+T-3}\rangle$. If the fusion process fails, the three W states $|W_{N-1}\rangle$, $|W_{M-1}\rangle$ and $|W_{T-1}\rangle$ can be re-fused by the same procedure.}\label{Fig2}
\end{figure}
As depicted in FIG.\ref{Fig2}, to fuse three W states $|W_N\rangle_a$, $|W_M\rangle_b$ and $|W_T\rangle_c$ of sizes $N, M, T$, respectively, one atom will be extracted from each entangled system and sent through the cavity, and these three atoms are marked by $1, 2$ and $3$, respectively. The detuned interaction between the three atoms and the cavity mode will lead the following state evolution:
\begin{equation}
\begin{aligned}
|W_N\rangle_a&\otimes|W_M\rangle_b\otimes|W_T\rangle_c=|(N-1)_g,e\rangle/\sqrt{N}\otimes|(M-1)_g,e\rangle/\sqrt{M}\otimes|(T-1)_g,e\rangle/\sqrt{T}\\
&\longrightarrow 1/\sqrt{NMT}|(N-1)_g\rangle_{\widetilde{1}}|(M-1)_g\rangle_{\widetilde{2}}|(T-1)_g\rangle_{\widetilde{3}}e^{-i3\lambda t}|e_1\rangle|e_2\rangle|e_3\rangle \\
&+1/\sqrt{NMT}|(N-1)_g\rangle_{\widetilde{1}}|(M-2)_g,e\rangle_{\widetilde{2}}|(T-1)_g\rangle_{\widetilde{3}}e^{-i\lambda t}(A|g_1\rangle|e_2\rangle|e_3\rangle+B|e_1\rangle|g_2\rangle|e_3\rangle+A|e_1\rangle|e_2\rangle|g_3\rangle)\\
&+1/\sqrt{NMT}|(N-2)_g,e\rangle_{\widetilde{1}}|(M-1)_g\rangle_{\widetilde{2}}|(T-1)_g\rangle_{\widetilde{3}}e^{-i\lambda t}(B|g_1\rangle|e_2\rangle|e_3\rangle+A|e_1\rangle|g_2\rangle|e_3\rangle+A|e_1\rangle|e_2\rangle|g_3\rangle)\\
&+1/\sqrt{NMT}|(N-2)_g,e\rangle_{\widetilde{1}}|(M-2)_g,e\rangle_{\widetilde{2}}|(T-1)_g\rangle_{\widetilde{3}}(B|g_1\rangle|g_2\rangle|e_3\rangle+A|g_1\rangle|e_2\rangle|g_3\rangle+A|e_1\rangle|g_2\rangle|g_3\rangle)\\
&+1/\sqrt{NMT}|(N-1)_g\rangle_{\widetilde{1}}|(M-1)_g\rangle_{\widetilde{2}}|(T-2)_g,e\rangle_{\widetilde{3}}e^{-i\lambda t}(A|g_1\rangle|e_2\rangle|e_3\rangle+A|e_1\rangle|g_2\rangle|e_3\rangle+B|e_1\rangle|e_2\rangle|g_3\rangle)\\
&+1/\sqrt{NMT}|(N-1)_g\rangle_{\widetilde{1}}|(M-2)_g,e\rangle_{\widetilde{2}}|(T-2)_g,e\rangle_{\widetilde{3}}(A|g_1\rangle|g_2\rangle|e_3\rangle+A|g_1\rangle|e_2\rangle|g_3\rangle+B|e_1\rangle|g_2\rangle|g_3\rangle)\\
&+1/\sqrt{NMT}|(N-2)_g,e\rangle_{\widetilde{1}}|(M-1)_g\rangle_{\widetilde{2}}|(T-2)_g,e\rangle_{\widetilde{3}}(A|g_1\rangle|g_2\rangle|e_3\rangle+B|g_1\rangle|e_2\rangle|g_3\rangle+A|e_1\rangle|g_2\rangle|g_3\rangle)\\
&+1/\sqrt{NMT}|(N-2)_g,e\rangle_{\widetilde{1}}|(M-2)_g,e\rangle_{\widetilde{2}}|(T-2)_g,e\rangle_{\widetilde{3}}|g_1\rangle|g_2\rangle|g_3\rangle.\label{eq10}
\end{aligned}
\end{equation}

After flying out of the cavity, atomic state measurement will be made on the three atoms. If two of the three atoms are detected in excited states, the remaining $(N+M+T-3)$ atoms will be left in $|W_{N+M+T-3}\rangle$ state. For instance, if the detection result is $|g_1\rangle|e_2\rangle|e_3\rangle$, the remaining atoms will be in the following state
\begin{eqnarray}
|\varphi_{gee}\rangle&=&\frac{e^{-i\lambda t}}{\sqrt{NMT}}[A|(N-1)_g\rangle_{\widetilde{1}}|(M-2)_g,e\rangle_{\widetilde{2}}|(T-1)_g\rangle_{\widetilde{3}}\nonumber\\
&+&B|(N-2)_g,e\rangle_{\widetilde{1}}|(M-1)_g\rangle_{\widetilde{2}}|(T-1)_g\rangle_{\widetilde{3}}+A|(N-1)_g\rangle_{\widetilde{1}}|(M-1)_g\rangle_{\widetilde{2}}|(T-2)_g,e\rangle_{\widetilde{3}}]\label{eq11}
\end{eqnarray}

By controlling the velocity of the atoms appropriately, $\lambda t=2\pi/9$ can be satisfied, and thus the state in Eq.(\ref{eq11}) will become:
\begin{eqnarray}
|\varphi_{gee}\rangle&=&\frac{1}{\sqrt{3NMT}}[|(N-1)_g\rangle_{\widetilde{1}}|(M-2)_g,e\rangle_{\widetilde{2}}|(T-1)_g\rangle_{\widetilde{3}}\nonumber\\
&+&e^{i2\pi/3}|(N-2)_g,e\rangle_{\widetilde{1}}|(M-1)_g\rangle_{\widetilde{2}}|(T-1)_g\rangle_{\widetilde{3}}+|(N-1)_g\rangle_{\widetilde{1}}|(M-1)_g\rangle_{\widetilde{2}}|(T-2)_g,e\rangle_{\widetilde{3}}],\label{eq12}
\end{eqnarray}
where a global phase factor $e^{-i19\pi/18}$ has been discarded.

Although the state in Eq.(\ref{eq12}) is not the standard W state:
\begin{eqnarray}\label{eq13}
|\varphi_{gee}\rangle=\frac{\sqrt{N+M+T-3}}{\sqrt{3NMT}}|W_{N+M+T-3}\rangle,
\end{eqnarray}
there is only a relative phase difference between the state and the standard W state, which can be removed by a classical pulse on any one of the $(N+M+T-3)$ atoms. The total success probability for the fusion process is
\begin{eqnarray}\label{eq14}
 P_{N+M+T-3}=\frac{(N+M+T-3)}{NMT}.
\end{eqnarray}

If the detection result is $|g_1\rangle|g_2\rangle|g_3\rangle$, the state of the remaining atoms becomes
\begin{eqnarray}
|\varphi_{ggg}\rangle &=&\frac{1}{\sqrt{NMT}}|(N-2)_g,e\rangle_{\widetilde{1}}|(M-2)_g,e\rangle_{\widetilde{2}}|(T-2)_g,e\rangle_{\widetilde{3}}\nonumber\\
&=&\frac{\sqrt{(N-1)(M-1)(T-1)}}{\sqrt{NMT}}|W_{N-1}\rangle|W_{M-1}\rangle|W_{T-1}\rangle, \label{eq15}
\end{eqnarray}
which is a product of three new W states $|W_{N-1}\rangle, |W_{M-1}\rangle, |W_{T-1}\rangle$, i.e. each state loses one atom. These three new W states can be re-fused again in the same way.

Now, it is necessary to have a look at the experimental feasibility of our scheme. If we use the Rydberg atomic levels with principal quantum numbers $49$, $50$ and $51$, the coupling strength can reach  $g=2\pi\times24 $ kHz \cite{brune}, and the atomic radiative time is around $T_r=3\times10^{-2} s$ \cite{osnaghi,yang1,zheng,guo}. In our scheme, we set $\delta=10g$, so the operation time for the fusion process is about $10^{-4}s$. In addition, the quality factor of order $Q=10^8$ has been realized for a cavity \cite{brune}, and the efficient decay time of this kind of cavity ($50$ GHz) is approximately $3\times10^{-2} s$. So, the fusion operation of our scheme can be realized before atomic decay and cavity decay, which means that the proposed schemes are feasible.

\section*{Discussion}

  We have presented two schemes to fuse W states in cavity QED system. In the first scheme, one auxiliary atom is introduced to fuse two W states $|W_N\rangle_a$ and $|W_M\rangle_b$ into a $(N+M-1)$-atom W state with probability $2(N+M-1)/3NM$. As a by-product, a $(N+M-2)$-atom W state can be generated if the main fusion process fails. In the second scheme, three atoms from three W states $|W_N\rangle_a$, $|W_M\rangle_b$ and $|W_T\rangle_c$, respectively, will interact with a detuned cavity mode, which can fuse the three W states into a $(N+M+T-3)$-atom W state with probability $(N+M+T-3)/NMT$. In both of these schemes, if no excitation is detected from the extracted atoms, i.e. the total fusion process fails, the remaining atoms are still in the product of two or three new W states, which can be re-fused by the same fusion process too. That is to say, the fusion process can work in an iterative manner, which will greatly reduce the entanglement waste during the process. The requirement of creating initial $W_3$ states disappears, i.e. W states of sizes $N=2, M=2$ can be fused as well. In addition, a comparison between our fusion schemes and the schemes of Refs \cite{ozaydin,yesilyurt,bugu} indicates that our schemes don't need a complicated Fredkin gate. In addition, the feasibility analysis indicates that our schemes maybe implementable within the current experimental technology.

\section*{Methods}
\subsection*{Detuning interaction between atoms and the cavity mode }

 The key step of our fusion mechanism is the detuned interaction between three identical atoms and a cavity mode, which can be described by the following interaction Hamiltonian(in the interaction picture)\cite{zheng}
\begin{equation}\label{eq16}
H_i=g\sum_{j=1,2}(e^{-i \delta t} a^\dag S^{-}_j+e^{i\delta t}a S^{+}_j),
\end{equation}
where $|e_{j}\rangle$ and $|g_{j}\rangle$ are the excited and ground states of the $j$th atom, respectively, and $S^+_j=|e_j\rangle\langle g_j|$, $S^-_j=|g_j\rangle\langle e_j|$. $a$ and $a^\dagger$ are the annihilation and creation operators for the cavity mode, respectively. $\omega_0$ is the atomic transition frequency, $\omega$ is the frequency of the cavity mode, and $\delta=\omega_0-\omega$ is the detuning between them. The coupling strengths between each atom and the cavity mode are supposed to be equal and described by the parameter $g$. The prerequisite of this scheme is that the interaction is far-off-resonant, so that there is no energy exchange between the atomic system and the cavity mode during the interaction. This condition can be satisfied if $\delta\gg g$, and then the effective Hamiltonian becomes\cite{guo}:

\begin{equation}\label{eq17}
H=\lambda\left[\sum^{n}_{i,j=1}(S^+_jS^-_iaa^\dag -S^-_jS^+_ia^\dag a)\right],
\end{equation}
where $\lambda=g^2/\delta$. For simplicity, we consider the case where the cavity mode is initially prepared in the vacuum state. For the case of $n=3$, the atom-cavity system effective Hamiltonian reduces to

\begin{equation}\label{eq18}
H_{eff}=\lambda\left(\sum^{3}_{j=1}|e\rangle_{jj}\langle e|+\sum^{3}_{i,j=1,i\neq j}S^+_jS^-_i\right).
\end{equation}

\section*{Acknowledgements}
This work is supported by the National Natural Science Foundation of China (NSFC) under Grants No.11274010, and No.11374085; the Specialized Research Fund for the Doctoral Program of Higher Education (Grants No.20113401110002); Anhui Provincial Natural Science Foundation No.1308085QA18; The Key Program of the Education Department of Anhui Province under Grants No.KJ2013A195 and No.KJ2013B172; the personnel department of Anhui Province. F. Ozaydin is funded by Isik University Scientific Research Funding Agency under Grant Number: BAP-14A101.

\section*{Author contributions statement}

X. P. Zang and M. Yang carried out the calculations. M. Yang and F. Ozaydin conceived the idea. All authors contributed to the interpretation of the results and the writing of the manuscript. All authors reviewed the manuscript.

\section*{Additional information}

\textbf{Competing financial interests:} The author declares no competing financial interests.

\end{document}